\documentclass[12pt,preprint]{aastex}

\usepackage{epsfig,rotating,amsmath,subfigure}
\usepackage{natbib,lscape}

\shorttitle{GALFA-HI Survey of Compact Clouds}
\shortauthors{Saul et al.}

\slugcomment{Accepted for Publication in The Astrophysical Journal}

\def\spose#1{\hbox to 0pt{#1\hss}}
\def\simlt{\mathrel{\spose{\lower 3pt\hbox{$\mathchar"218$}}
     \raise 2.0pt\hbox{$\mathchar"13C$}}}
\def\simgt{\mathrel{\spose{\lower 3pt\hbox{$\mathchar"218$}}
     \raise 2.0pt\hbox{$\mathchar"13E$}}}

\def\gtrapprox  {\;\lower 0.5ex\hbox{$\buildrel >\over \sim\ $}}
\def\lessapprox {\;\lower 0.5ex\hbox{$\buildrel < \over \sim\ $}}

\def\deg        {$^\circ$}

\def\kms        {km~s$^{-1}$}

\def\vlsr	{V$_{\rm LSR}$}
\def\vgsr   {V$_{\rm GSR}$}

\begin{document}

\title{The GALFA-HI Compact Cloud Catalog}

\author{Destry R. Saul\altaffilmark{1}, J. E. G. Peek\altaffilmark{1,2}, J. Grcevich\altaffilmark{1}, M. E. Putman\altaffilmark{1}, K. A. Douglas\altaffilmark{3}, E. J. Korpela\altaffilmark{4}, S. Stanimirovi\'{c}\altaffilmark{5}, C. Heiles\altaffilmark{7}, S. J. Gibson\altaffilmark{8}, M. Lee\altaffilmark{5}, A. Begum\altaffilmark{6}, A. R. H. Brown\altaffilmark{2}, B. Burkhart\altaffilmark{5}, E. T. Hamden\altaffilmark{2}, N. M. Pingel\altaffilmark{5}, S. Tonnesen\altaffilmark{9}}

\altaffiltext{1}{Department of Astronomy, Columbia University, New York, NY 10027}
\altaffiltext{2}{Hubble Fellow}
\altaffiltext{3}{University of Calgary/Dominion Radio Astrophysical Observatory, P.O. Box 248, Penticton, BC V2A 6J9, Canada}
\altaffiltext{4}{Space Sciences Laboratory, University of California, Berkeley, CA 94720}
\altaffiltext{5}{University of Wisconsin, Madison, 475 N Charter St, Madison, WI 53703}
\altaffiltext{6}{Indian Institute of Science Education and Research, ITI Campus (Gas Rahat) Building, Govindpura, Bhopal - 23, India}
\altaffiltext{7}{Radio Astronomy Lab, UC Berkeley, 601 Campbell Hall, Berkeley, CA 94720}
\altaffiltext{8}{Department of Physics and Astronomy, Western Kentucky University, Bowling Green, KY 42101}
\altaffiltext{9}{Department of Astrophysical Sciences, Princeton University, Princeton, NJ 08544}

\begin{abstract}
We present a catalog of 1964 isolated, compact neutral hydrogen clouds from the Galactic Arecibo L-Band Feed Array Survey Data Release One (GALFA-HI DR1). 
The clouds were identified by a custom machine-vision algorithm utilizing Difference of Gaussian kernels to search for 
clouds smaller than 20\arcmin. The clouds have 
velocities typically between $|$\vlsr$|$ $= 20 - 400$ \kms, linewidths of $2.5-35$ \kms, and column densities ranging from $1-35 \times 10^{18}$ cm$^{-2}$.
The distances to the clouds in this catalog may cover several orders of magnitude, so the masses may range from less than a Solar mass for clouds within the Galactic disc, to greater than $10^4$ M$_\odot$ for HVCs at the tip of the Magellanic Stream.
To search for trends, we separate the catalog into five populations based on position, velocity, and linewidth: 
high velocity clouds (HVCs); galaxy candidates; cold low velocity clouds (LVCs); 
warm, low positive-velocity clouds in the third Galactic Quadrant; and the remaining warm LVCs. 
The observed HVCs are found to be associated with previously-identified HVC complexes. 
We do not observe a large population of isolated clouds at high velocities as some models predict. 
We see evidence for distinct histories at low velocities in detecting populations of clouds corotating with the 
Galactic disc and a set of clouds that is not corotating.
\end{abstract}

%change as appropriate

\section{Introduction}

The recent discovery of compact clouds over the entire velocity range of our Galaxy and the Local Group has posed interesting questions as to their
role in galaxy evolution and the Galactic ISM \citep[e.g.][]{lockman02, ford08, ryanweber08, benbekhti09, stanimirovic08, heitsch09, begum10, giovanelli10, hsu11}. 
Despite the similarity of the individual clouds when observed with the hyperfine neutral hydrogen 21 cm line, different classes of clouds have been identified and many possible origins have been proposed.
The detection of stellar outflows in HI points to the possibility that some small clouds could be produced by evolved stars \citep[e.g.][]{matthews11}. At the disc-halo interface, a population of clouds has been observed rotating with the Galaxy that may have been pushed off the disc by stellar feedback
\citep{ford08, ford10}.
In the Galactic halo, compact clouds may represent the initial cooling seeds from the multiphase halo medium that will eventually fuel the disk \citep{maller04, joung11}, or the remnants of larger complexes \citep{heitsch09}.  
The continual production of these clouds could provide a significant source of the fuel for the Milky Way's ongoing star formation. 
The abundance of newly discovered small Local Group dwarf galaxies \citep[and others]{belokurov2006,zucker2006,irwin07}, one of which is particularly gas-rich \citep{ryanweber08}, raises the possibility that some of these clouds represent previously undiscovered dwarf galaxies. 
It is even possible that some compact clouds are outside the halo and physically large \citep{giovanelli10}.

Unraveling the nature of the compact clouds requires a large, well-defined sample. Historically, finding such a sample has been difficult due to the limited area, sensitivity, and resolution covered by HI surveys,
as well as the difficulty of recovering compact objects confused by diffuse Galactic emission \citep{stil06, ford08}.    The latter problem is evident in the lack of detections of halo clouds (commonly referred to as high-velocity clouds; HVCs) between \vlsr $= -90$ to 90 \kms \citep{wakker97,peek09}, and the recent, optical discovery of gas-rich dwarf satellites at velocities where Galactic emission is prevalent \citep{irwin07}.   
A large, sensitive survey with sufficient kinematic and spatial resolution to recognize clouds that are near Galactic emission is needed.

The GALFA-HI (Galactic Arecibo L-Band Feed Array HI) Survey \citep{peek11} provides the data needed to detect a large sample of compact clouds across the range of velocities of the Galaxy and Local Group. At completion it will cover 13,000 deg$^2$ (58\% of which is available in Data Release 1\footnotemark and cataloged here; DR1)
\footnotetext{The GALFA-HI DR1 is publicly available at http://purcell.ssl.berekley.edu}
over -1\deg $< \delta <$ 38\deg~and the entire RA range between \vlsr $=\pm650$ \kms. The GALFA-HI survey has a spatial resolution of 4\arcmin~and a channel spacing of 0.184 \kms. The sensitivity at 6$\sigma$ to compact, warm (FWHM = 15 \kms) HI clouds corresponds to masses as low as 10$^{-2}$ M$_\odot$ at 1 kpc and 10$^{4}$ M$_\odot$ at 1 Mpc (see \S \ref{sec:sense}). As the sensitivity of a survey to compact, spectrally resolved sources scales inversely with both the antenna beam solid angle and the noise for a given channel width, GALFA-HI is uniquely positioned among other large-area Galactic HI surveys. With the sensitivity of a single-dish radio telescope and resolution approaching that of a compact array, GALFA-HI is 2.5 to 80 times more sensitive to compact clouds than other large-area surveys (2.5: IGPS; \citealt{taylor03,stil06b,mccluregriffiths05}, 15:GASS/ EBHIS; \citealt{mccluregriffiths09,winkel10}, 80: LAB; \citealt{kalberla05}). Initial investigations of GALFA-HI data identified many compact clouds, both visually \citep{begum10, stanimirovic06} and through an automated algorithm \citep{hsu11}. 

In this paper we present a catalog of compact clouds (4\arcmin--20\arcmin) from the GALFA-HI Survey DR1 that has been created using a new technique designed to recover faint clouds and identify structures near diffuse Galactic emission. The methods we use are detailed in \S\ref{sec:method}, with the completeness of the catalog assessed in \S \ref{sec:sense} and the catalog presented in \S\ref{sec:properties}.   The properties of the clouds cataloged are described in \S \ref{sec:results} where the clouds are separated into five populations.  A discussion of the implications of our results can be found in \S\ref{sec:discussion}.

\section{Observations and Data Reduction}
\label{sec:obs}

\begin{figure}
\begin{center}
\includegraphics[scale=.6, angle=0]{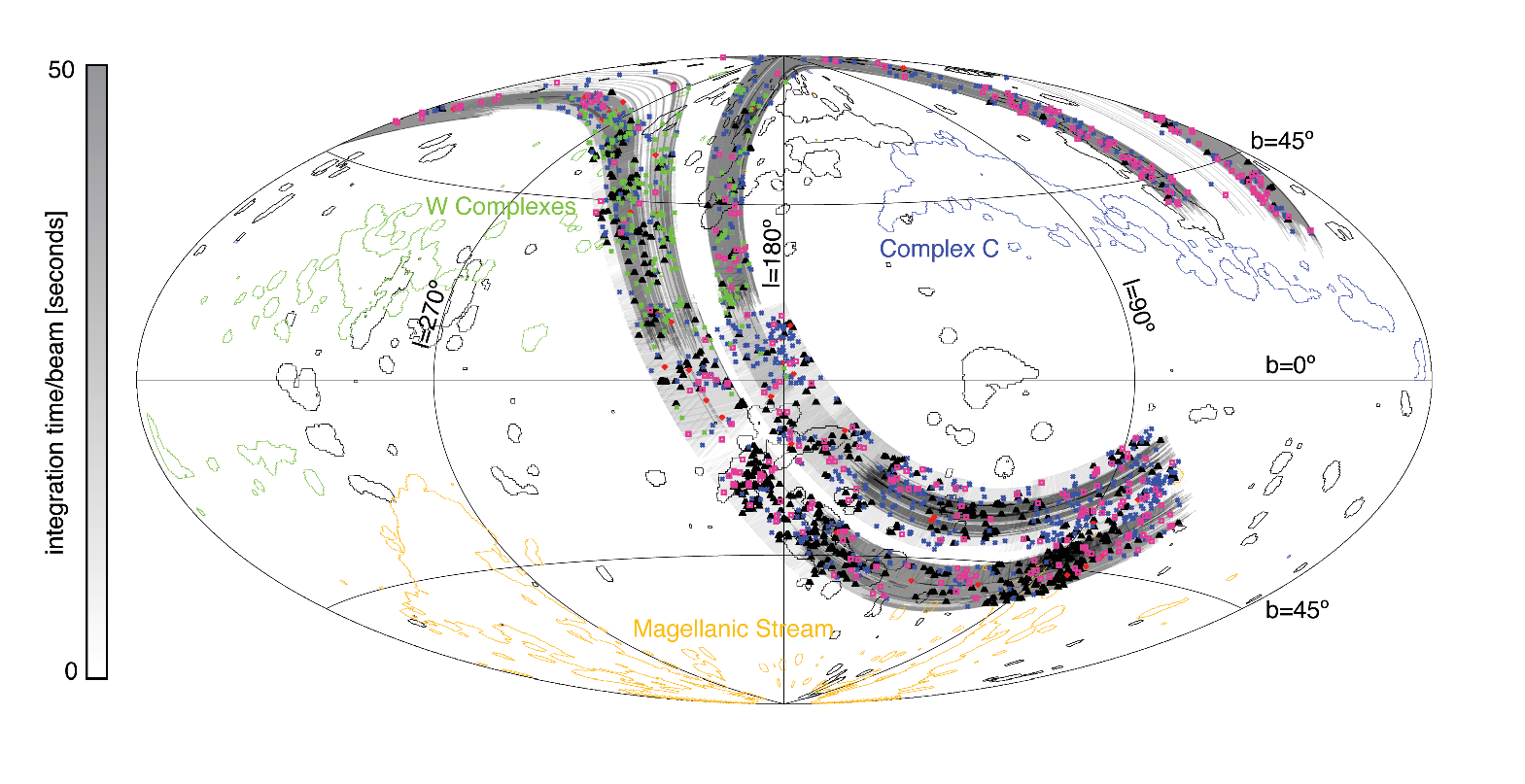}
\caption{A Hammer-Aitoff plot of the positions of all of the cataloged clouds in
Galactic coordinates. Cloud type is indicated by color, with black for HVCs, 
red for galaxy candidates, blue for cold LVCs, pink for warm LVCs (excluding the Q3 clouds), and green for
warm, positive-velocity Q3 clouds. The coverage of the survey is indicated in
grayscale, with the darkest gray indicating $> 50$ seconds per beam of
integration time. High velocity cloud complexes are shown in outline,
with the W complexes shown in green, complex C shown in blue, and the
Magellanic Stream shown in orange. Other HVCs are shown in black.
\label{fig:aitoff}}
\end{center}
\end{figure}

The catalog is generated using the GALFA-HI Survey DR1. GALFA-HI is a survey of the 1420 MHz hyperfine transition of neutral hydrogen in the Galaxy using the Arecibo 305m telescope and the ALFA seven-beam feed array.  The survey is completed commensally with other Arecibo extragalactic and Galactic surveys \citep{giovanelli05,guram09}.
GALFA-HI data provide a channel spacing of 0.184 \kms~and cover a velocity range of $\pm650$ \kms~in the Local Standard of Rest (LSR) with a spatial resolution of 4\arcmin.  The DR1 data cover 7520 square degrees of sky in an area between  $\delta = 38 $\deg~ and $\delta = - 1$\deg~ (see bottom panel of Figure \ref{fig:ra_dec}, and Figure \ref{fig:aitoff}), with a range of sensitivity from 120~mK to 50~mK in 0.74 \kms~channels.
The data were corrected for the effects of 
bandpass, gain, baseline ripple, and the first sidelobes of the ALFA beams with a special-purpose data reduction pipeline. The data are not stray radiation corrected. As the Arecibo stray sidelobes are large and low gain, the effect of stray radiation is to add a spurious low-amplitude HI signal that varies slowly as a function of position on the sky \citep{peek11}. Were Arecibo to have angularly compact stray sidelobes, broadband surveys such as GALFACTS \citep{guram09} would detect erroneous replications of tremendously bright continuum sources, an effect that has not been observed. The spatially broad contamination from stray radiation should not significantly affect our catalog of compact clouds. The details of GALFA-HI observing and data reduction, along with the specifics of the DR1 data set, can be found in \citet{peek11}.

\section{Cataloging Method}
\label{sec:method}

Identifying clouds by eye, as has been done in the past, is impractical for the amount of data
the GALFA-HI survey is producing. This, coupled with a desire for completeness over a broad range of velocities and 
sensitivities, motivated our development of an algorithm to identify and characterize compact isolated clouds in the GALFA-HI data cubes. 
The algorithm consists of four distinct components: Galactic Subtraction, Convolution, Region of Interest (ROI) Detection,
and Merge and Candidate Selection; each discussed in detail in \S \ref{sec:truffles}. 
After the initial detection, each cloud candidate was inspected to remove systematic noise spikes, RFI, and connected structures.
This parsing process is discussed in \S \ref{sec:classify}. In \S \ref{sec:sense} we show that our sensitivity is dominated by systematic errors
and that our sensitivity is effectively constant over our search space. The properties we measure for each cloud are explained in \S \ref{sec:properties}.

\subsection{Cloud Detection Algorithm}
\label{sec:truffles}
\noindent {\bf 1.  Galactic Subtraction} - To remove smooth Galactic emission,
we applied a high-pass spatial filter by smoothing each channel with a one square degree median box and subtracting the resulting smoothed cube
from the data. The intent of this pre-processing step is to remove emission on larger scales than our targets (20\arcmin) without removing signal
from the compact clouds. We chose a box size of one degree because an isolated 20\arcmin~cloud on a smooth background is unaffected by the filtering.
We applied the filter over the full velocity range to avoid introducing discontinuities in the data.
Filtering allows us to probe for compact clouds down to velocities closer to \vlsr$=0$~\kms~ than would otherwise be possible. We do not make a lower cut in velocity. Instead, we found the amount subtracted in this step
to be a good metric for cloud believability. We discuss this further in \S \ref{sec:classify}.

\begin{figure}
\includegraphics[scale=1.0,angle=0]{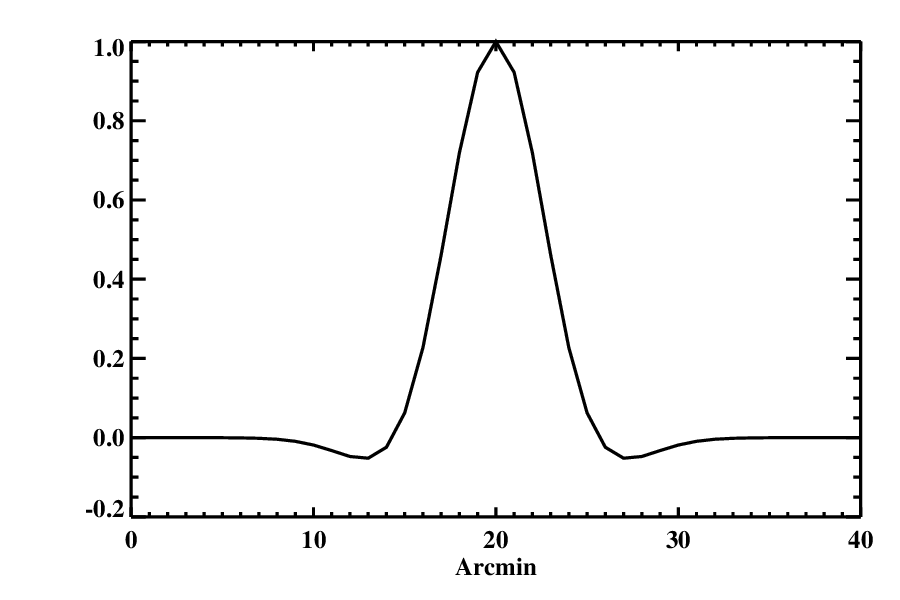}
\caption{An example of the Difference of Gaussians kernels used in the convolution step of our algorithm. This is a two dimensional slice through one of the three dimensional kernels. Here, the positive Gaussian has a FWHM of 7', while the negative Gaussian has a FWHM of 8.4'. The Gaussians are scaled such that the integral of the 3D kernel is zero.}
\label{fig:mexhat}
\end{figure}

\noindent{\bf 2.  Convolution} - To increase our sensitivity, we convolved the data with a three dimensional kernel in a simplified wavelet analysis.   Using this method enhances regions of isolated emission according to their similarity to the 3D kernels used.   We chose to use four ``Difference of Gaussians'' (DOG) kernels, one of which is shown in Figure~\ref{fig:mexhat}. The DOG kernel acts like a bandpass
filter and is computationally faster than other kernels since it can be separated into 1D and 2D kernels used in series \citep[See][and references therein]{sonka07}.
The kernel is the sum of two three-dimensional Gaussians: a positive Gaussian and a negative Gaussian that is wider in all dimensions by a factor of 1.2. 
The Gaussians are scaled such that the integral of the kernel is zero. We define the size of the kernel as the FWHM of the positive Gaussian.  

\begin{figure}
\includegraphics[scale=1.0,angle=0]{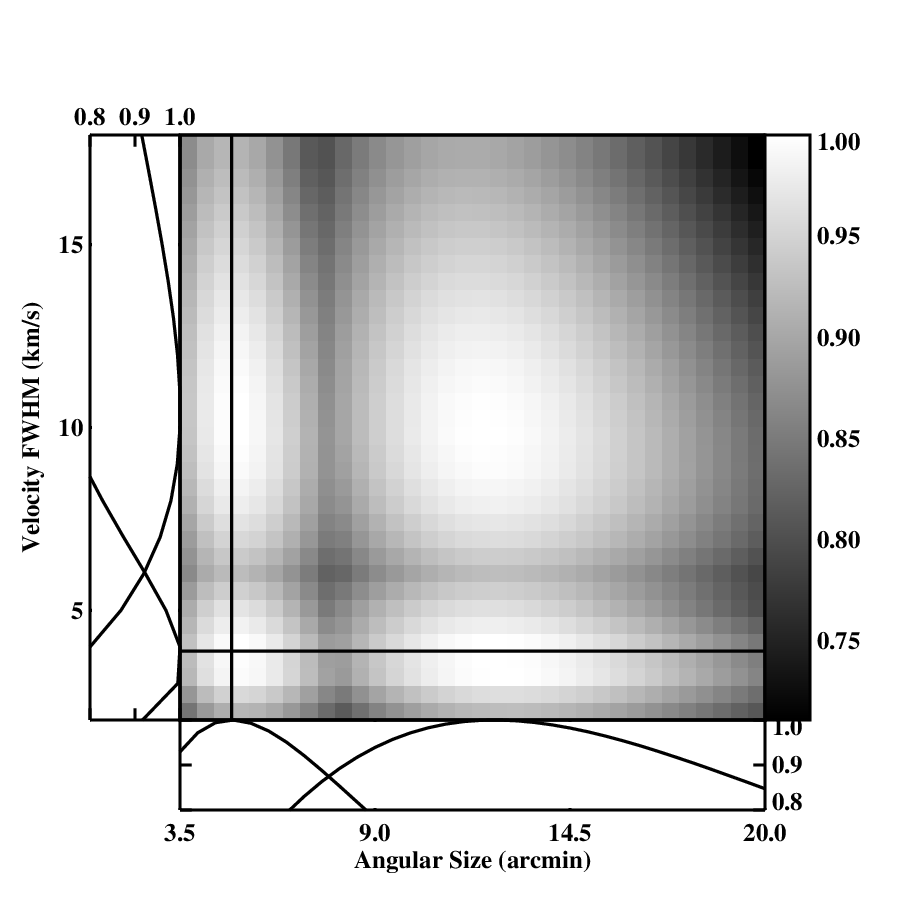}
\caption{The sensitivity to recovering clouds of given sizes and velocity widths given our chosen optimal kernels.  The cut-off at the lower end in size and velocity is set by the resolution of the data. Excluding the clouds that are both largest (20') and widest (20~\kms), where the sensitivity drops by 70\%, the sensitivity drops to no less than 80\% in the parameter space shown.}
\label{fig:kernels}
\end{figure}

We tuned the sizes and widths of the four kernels by measuring the peak response to a set of model clouds of fixed signal to noise.  The model clouds covered the range of possible observed cloud sizes and velocity widths given the data and chosen upper limits. The response of the optimum kernels is plotted in Figure~\ref{fig:kernels}. The four kernels are circular with FWHM and velocity widths of 7\arcmin~and 5~\kms, 7\arcmin~and 15~\kms, 18\arcmin~and 5~\kms, and 18\arcmin~and 15~\kms. The theoretical response is above 80\% for our entire search space except for clouds that are both large (20\arcmin) and wide (20~\kms) where the response drops to 70\%. While increasing the number of kernels would provide a 
smoother sensitivity function, the computational requirements are unreasonable.
\S \ref{sec:sense} discusses sensitivity further.

This step results in four convolved data cubes that will be passed to the region detecting routines. Before this, a careful treatment of noise is necessary to accurately identify levels of significance and to make it possible to compare the results for the four kernels.  The noise characteristics are complicated due to different regions of DR1 having varying amounts of data taken with different observing modes (see Figure~\ref{fig:ra_dec} and \cite{peek11} for details).  Also, the noise in each voxel is not statistically independent.  While the GALFA-HI data cubes are gridded to 1\arcmin~pixels, Arecibo's resolution is 4\arcmin. We also observe correlated noise on the scale of ALFA's beam pattern of $\sim$25\arcmin. In order to calculate the uncertainty in the convolved cubes we take the following steps. (1) We begin by taking the original data and finding the standard deviation along each line of sight over a velocity range that does not include significant emission.  This allows us to construct a noise map for the original data.  (2) Assuming the the value of the noise is the same over the full velocity range for each pixel, the noise detected by each kernel is calculated by multiplying the noise map by each kernel at each position, and then summing the scaled expected noise values in quadrature.  This creates a convolved noise map for each kernel.  (3) Dividing each channel in the convolved cubes by the corresponding convolved noise map converts the cube to units of signal to noise, for uncorrelated noise. (4) We normalize the data to correct for the effect of correlated noise in the unconvolved datacubes. We fit a Gaussian to the distribution of values in a region with no detectable emission. For uncorrelated noise, the standard deviation should be one. Dividing each convolved cube with the corresponding fit's standard deviation converts the units to signal to noise ($\sigma$).

\noindent{\bf 3. Region of Interest (ROI) Detection - }  We now have four convolved data cubes in units of $\sigma$ that can be used to detect regions of interest (ROI).  We use a simple `watertable' technique in which the highest peak in a cube is identified and the adjacent voxels above a watertable of 4$\sigma$ 
are added to create the ROI.  
The ROI is then numbered, saved, and deleted from the cube.  This is then repeated with the new highest peak in the cube selected and the ROI grown to the watertable level again.  The ROI detection process is completed when there are no peaks above a minimum of 6$\sigma$.  

\noindent{\bf 4. Merge and Candidate Selection- }  Having defined ROIs in all cubes corresponding to the four kernels, we need to merge the information to select clouds.   For this catalog we have chosen to include clouds with a spatial FWHM less than $20$\arcmin.  Our first step is to reject those clouds larger than this selected by our kernels.
For an ideal, Gaussian-shaped cloud with a spatial FWHM of 20\arcmin~and a peak brightness of 1K in the unconvolved data, 
the 7\arcmin~kernels would identify a ROI 34\arcmin~across and
the 18\arcmin~kernels would identify a ROI 36\arcmin~across.   Therefore ROIs larger than these values are rejected as extended, and any ROI's found within these larger structures in position-velocity space are flagged as ``embedded'' and also removed.  
We also removed ROIs that extend into velocities where the bandpass response begins to decline ($|$\vlsr$|> 650$ \kms).

We use four kernels of different sizes to achieve smooth sensitivity over the parameter space being searched.
And while the weakest clouds are only detected by the most similar kernel, brighter clouds can be detected by multiple kernels.
Therefore, the next part of the Merge and Candidate Selection step is to take the clouds found by multiple kernels and select the ROI for the cloud that has the highest peak value.  A higher response from a specific kernel indicates a better fit to the cloud.  The final list of $\sim$10,000 ROIs were visually inspected and classified as described in the next section.

\subsection{Cloud Classification}
\label{sec:classify}

The cloud detection algorithm is very effective at detecting significant isolated structures, but the ISM includes extremely complicated features.
At low velocities in particular, the Galactic emission is a web of needles, compact clouds at the end of filaments (peninsulas), bubbles, arcs, and many other shapes \citep[][and references therein]{kalberla09}.
Some of these structures were included in our candidate list, as well as some RFI and calibration artifacts.
Removing connected structures and artifacts requires visual inspection of the candidates.  When visual inspection began it was soon realized that those ROIs in regions where more than 1K was removed in the Galactic Subtraction step were generally part of these more extended structures.   We therefore rejected those ROIs, resulting in the gap in velocity coverage seen in Figures \ref{fig:ra_dec}, \ref{fig:line_v_plot}, and \ref{fig:ra_v_hvc} (generally $|$\vlsr$|< 20$ \kms).

The remaining ROIs were examined visually to determine if they were part of a more extended Galactic structure, RFI or a calibration artifact, and should be removed.  
To be included in the final catalog, we required that a cloud was visually inspected by at least three of the authors, with a majority in agreement.
Since the clouds are detected by integrating three dimensional volumes, visual inspection is a non-trivial process. With thousands of clouds to inspect and ten viewers, bookkeeping was also non-trivial. To solve both of these problems, we built a procedure which displays integrated position-position images and average spectra similar to those in Figure~\ref{fig:cat}.  If this information was not enough to classify the ROI, the
user could scan through the channel images and spectra of the ROI in the original or Galactic-subtracted cubes. 
The viewer then classifies the ROI as a real discrete cloud, a cloud that is part of a more extended structure, or a glitch in the data. 
Only ROIs in the first category are included in the final cloud catalog.  A ROI was considered part of a more extended structure and excluded from the catalog if there was emission $>4\sigma$ connecting the ROI to other areas of emission.   This is 4$\sigma$ in the unconvolved cube and represents structures that are not well enough defined by the kernels to appear significant in the convolved cubes.
If there are other areas of emission nearby, either in the form of other compact clouds or a larger complex, the cloud was included as long as it was not connected at the 4$\sigma$ level. Clouds with average spectra so weak that the fitting described in \S\ref{sec:properties} failed were also removed from the catalog.

After pruning the catalog, the remaining clouds were compared to the NED database to remove any previously known galaxies.  We searched for known objects within 50~\kms~in velocity and 10\arcmin~in position for each of the catalog candidates. Those candidates in agreement with known galaxies were removed from the catalog. Of the 2025 candidate clouds after visual evaluation, 61 were found to be associated with known galaxies and were removed leaving a final catalog of 1964 clouds. 

\subsection{Empirical sensitivity}
\label{sec:sense}

To fully understand the cloud sample we must understand the limitations and sensitivity of the cataloging process. Since the background contaminants can deviate dramatically from Gaussian noise and our cataloging method is non-linear, it would be very difficult to accurately determine our sensitivity only using an analytic technique. As such, we test our sensitivity using a Monte Carlo method; we inject false clouds into the data set, run our entire cataloging process and find which of these clouds we can recover. The false clouds we inject are 3D Gaussians, with variable RA, Dec., velocity,  peak brightness (in Kelvin; $T_{pk}$), major axis (FWHM in arcminutes; $a$), aspect ratio ($b/a$), linewidth (FWHM in \kms; $\Delta$V) and position angle (in degrees; $\phi$). We inject them into two regions, one with integration times ($t$) typically larger than 40 seconds per beam, and one with integration times typically smaller than 40 seconds per beam. We note here that many of the real clouds we detect in our catalog vary from simple Gaussian shapes, which will typically lower our sensitivity for a given set of parameters. 

\begin{figure*}
\begin{center}
\includegraphics[scale=.78, angle=0]{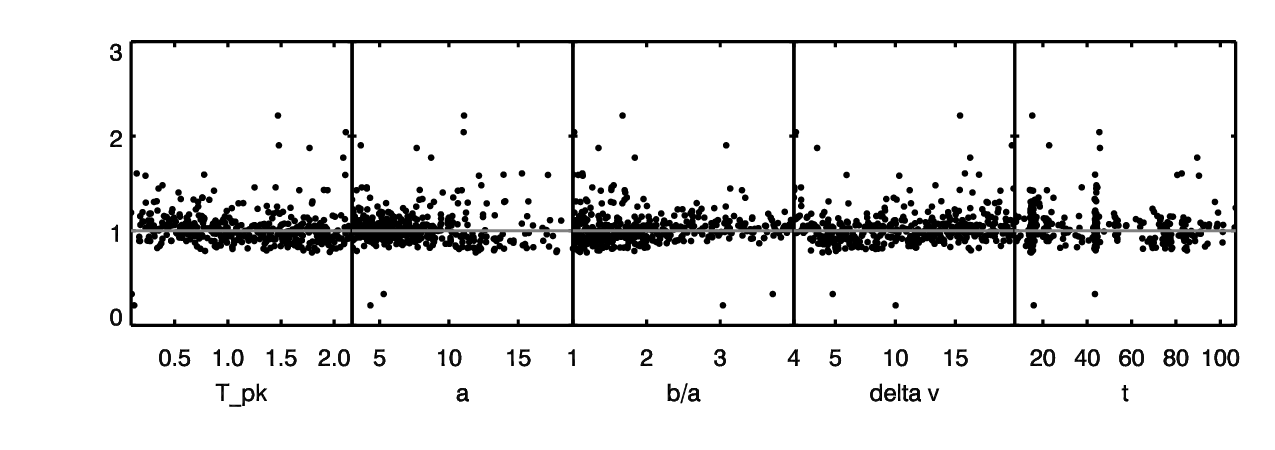}
\caption{The ratio of our sensitivity function, $S\left(T_{pk},a, b/a, \delta v, t\right)$, to the signal detected in the convolved data, $\sigma_{cl}$, as a function of various parameters of the injected clouds. We find we can accurately predict $\sigma_{cl}$ using Equation \ref{eq:Sfn_eq}. }
\label{fig:Sfn}
\end{center}
\end{figure*}

We find that our detection fraction is entirely independent of position angle. In position and velocity, clouds are more difficult to detect when they are near the edge of the mapped area or in areas of strong emission where the Galactic Subtraction step removed more that 1K  (See \S\ref{sec:truffles}). 
Ignoring for the moment the effect of background HI on our sensitivity, we use the remaining parameters ($T_{pk}$, $a$, $b/a$, $\Delta {\rm V}$, and $t$), to construct a metric, $S\left(T_{pk},a, b/a, \Delta {\rm V}, t\right)$, to quantify the extent to which a cloud is detected. Since by construction we do not detect clouds whose value in the convolved cubes, $\sigma_{\rm MAX}$, is below 6$\sigma$, we fit the functional form $S\left(T_{pk},a, b/a, \Delta {\rm V}, t\right)$ to match $\sigma_{\rm MAX}$, so as to best predict whether $\sigma_{\rm MAX} \ge 6\sigma$ and thus be detected. We find that a simple exponential scaling,
\begin{equation}\label{eq:Sfn_eq}
S\left(T_{pk},a, b/a, \Delta {\rm V}, t\right) = AT_{pk}^{\alpha}a^{\beta} (b/a)^{\gamma}\Delta {\rm V}^{\delta}t^{\epsilon},
\end{equation}
rather accurately captures the sensitivity function. We find $A = 0.86 \sigma, \alpha = 0.94, \beta = 0.57, \gamma = 0.18, \delta =0.30, \epsilon=0.50$, using a least-squares fit, minimizing the difference between $\sigma_{\rm MAX}$ and $S$. In Figure \ref{fig:Sfn} we plot $S\left(T_{pk},a, b/a, \Delta {\rm V}, t\right)/\sigma_{\rm MAX}$ against each of the parameters, to demonstrate that there is very little functional dependence of $S/\sigma_{\rm MAX}$ to any of these parameters. We note that at values of $a$ and $\Delta$V below the size of our smallest kernel (5', 7~\kms) $\sigma_{\rm MAX}$ decreases somewhat more quickly than the simple functional form in \ref{eq:Sfn_eq}, although the change is less than 50\%. We find that we detect 90\% of the clouds with $S > 6 \sigma$. The 10\% we do not detect are typically coincident with other HI structures or glitches in the data, and thus not detected as compact. 
We found that the fraction of false clouds we expect to detect that go undetected is independent of background brightness, once we ignore clouds with background brightnesses above 1K. This confirms that our background brightness cut of 1K is relatively conservative.

Comparisons to previous catalogs provide another metric for our detection algorithm. We compare our catalog to the work of \cite{begum10} and
\cite{hsu11}. The clouds of the Begum et al. catalog were identified visually in the high-resolution GALFA-HI data cubes. They found 96 clouds with all
but two at $|{\rm V_{LSR}}| < 90$~\kms. We include 31 of the Begum et al. clouds in our catalog. Of the remaining 65 clouds, 64 clouds were never
selected as candidates due to being connected to extended emission. 
The one cloud that was selected as a candidate, but not included in our catalog,
was removed because it was connected to the edge of an unobserved region. 
The median $|{\rm V_{LSR}}|$ of the undetected Begum et al.~clouds is 31 \kms~since many of these clouds are near Galactic emission.
The positions, velocities, linewidths, 	and sizes of clouds in both catalogs are in agreement when the 
different methods used are considered. The measured velocities agree within 1 \kms~and the typical 
size difference is less than 2\arcmin. Begum et al.~use an average spectrum to measure brightness temperature and linewidth. They therefore measure 
lower peak brightnesses, and the typical difference in measured linewidth is less than 3 \kms.

We identify 26 of the Magellanic Stream clouds found by \cite{hsu11}. Of the remaining Hsu et al.~Stream clouds, 19 are larger than 30\arcmin~and are excluded, 15 
were detected but were connected to other structures, and 15 were below our signal-to-noise threshold. The velocities and linewidths measured by
Hsu et al. are taken from an average spectrum over the region identified by the Duchamp detection algorithm \citep{whiting08}. Duchamp identifies 
regions over a given threshold (3$\sigma$ for Hsu et al.)  which are larger than the $4\sigma$ regions we identify. This introduces differences in the measured
velocities and line widths of typically 5 \kms. The brightness temperatures are measured from the raw data and agree with our measurements 
to within 0.1 K. These results indicate that our algorithm is effective at identifying and characterizing clouds that meet our criteria: smaller than 20\arcmin, 
brighter than 6$\sigma$ in the convolved data, and separated from complex, extended structure in position and velocity. 

\subsection{Cloud Properties}
\label{sec:properties}

\begin{figure}
\subfigure{\epsfig{file=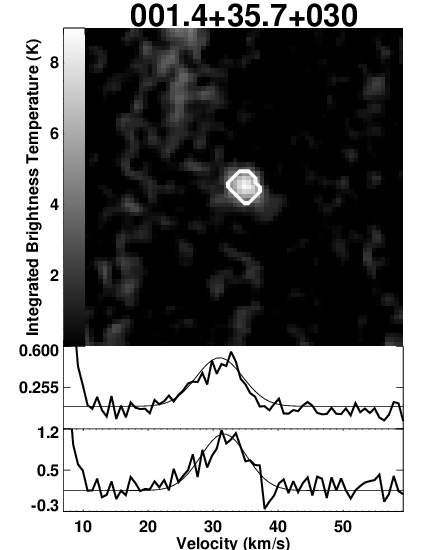,width=2in}}
\subfigure{\epsfig{file=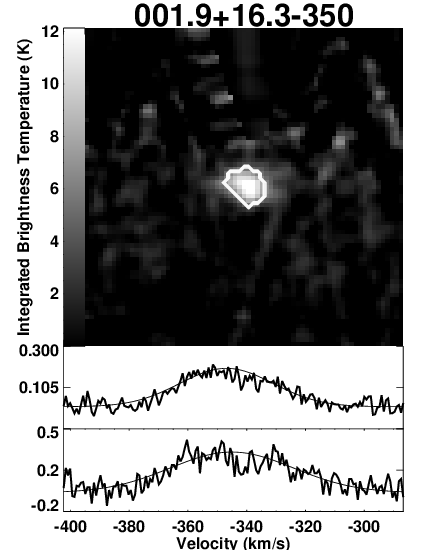,width=2in}}
\subfigure{\epsfig{file=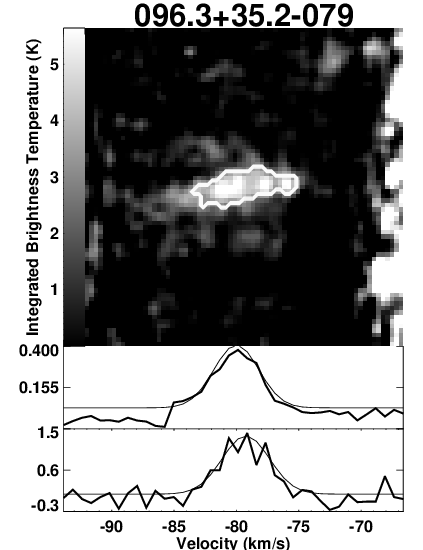,width=2in}}

\subfigure{\epsfig{file=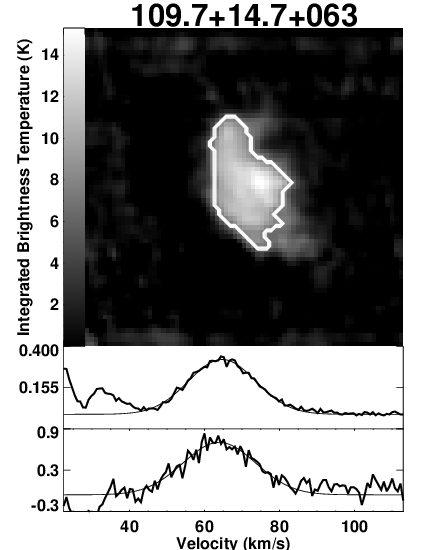,width=2in}}
\subfigure{\epsfig{file=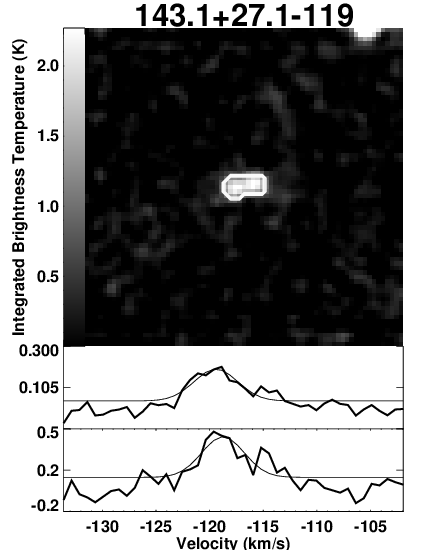,width=2in}}
\subfigure{\epsfig{file=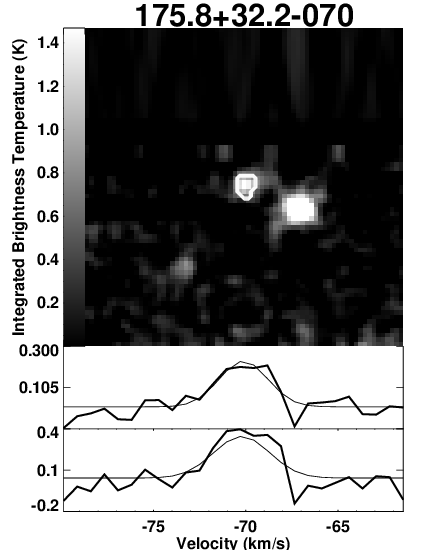,width=2in}}

\caption{A random sample of the DR1 GALFA-HI cloud images and spectra (the full version is available online). The image is produced by scaling each 
velocity channel by the result of the fit to the average spectrum (upper spectrum) and then integrating along velocity. This de-emphasizes nearby 
emission at velocities away from the peak of the cloud while still increasing the fidelity of the image over an unintegrated slice. The contour follows the half-maximum
of contiguous emission surrounding the peak position and is the reported area of the cloud.
The upper spectrum is the average over the Region of Interest with the four-parameter Gaussian fit (see \S\ref{sec:truffles} and \S\ref{sec:properties}).
The lower spectrum is the spectrum through the peak in the unconvolved data. If a Gaussian fit was successful on this spectrum, it is plotted. The full catalog is available at http://sites.google.com/site/galfahiccc.
\label{fig:cat}}
\end{figure}

To characterize the 1964 identified clouds, we began by fitting a four parameter Gaussian to the average spectrum over the ROI. The four parameters were
amplitude, velocity, linewidth, and a constant baseline. Using the results of the fit, we then refined the velocity range and fit the average spectrum over the new velocity range with the same single, four parameter Gaussian. 
The images in the catalog (see Figure \ref{fig:cat} for an excerpt) were prepared by scaling each channel by the 
normalized average fit and then integrating over velocity. This produces cleaner images and deemphasizes nearby structures at velocities where the 
clouds of interest are weakest.
To calculate the maximum column density of the cloud, we fit the same type of single Gaussian to 
the location of the brightest peak in the integrated image. When the fit for the maximum column density fails we report the average column density as a lower limit. 

A sample of the catalog is presented in Table~\ref{tab:cattab} and Figure \ref{fig:cat}, while the online version contains the complete catalog.\footnotemark ~The definitions of the 
cloud characteristics listed in Table~\ref{tab:cattab} are summarized below.   The parameters obtained from the convolved cubes are noted and the others are obtained from the original cubes after Galactic subtraction.

\footnotetext{The online materials are publicly available at http://sites.google.com/site/galfahiccc}

\begin{itemize}

\item{Cloud ID - Clouds are named by the right ascension (in degrees), declination, and \vlsr~velocity of the peak in the convolved cube.}

\item{RA \& DEC -  The right ascension and declination of the peak in the convolved cube. Note that this is not necessarily the location of peak flux, or the spatial center of the region of interest (see \S\ref{sec:truffles}).}

\item{$l$ \& $b$ - The Galactic longitude and latitude corresponding to the right ascension and declination.}

\item{$\sigma_{\rm MAX}$ - The maximum signal-to-noise of the cloud in the convolved cube. This is a measure of the significance of the cloud.}

\item{Size - The angular FWHM assuming a circular cloud, $2\sqrt{\rm Area/\pi}$, where the area is that of the half maximum value contour shown in the catalog images (Figure~\ref{fig:cat}). Due to our size constraint of 20\arcmin~along any axis, the selection function for elongated clouds drops below 20\arcmin. }

\item{V$_{\rm LSR}$ - The Local Standard of Rest velocity of the peak of the fit to the average spectrum. }

\item{V$_{\rm GSR}$ - The corresponding Galactic Standard of Rest velocity defined by \\ \vgsr=\vlsr+220sin($l$)cos($b$).}

\item{$\Delta V$ - The FWHM in velocity of the fit to the average spectrum.}

\item{T$_{Pk}$ - The peak brightness temperature after Galactic subtraction.}

\item{N$_{\rm HI}$ - The maximum HI column density calculated using the fit of the spectrum through the peak of the cloud in the unconvolved data. Where the fitting failed on the maximum spectrum, the average column density is reported as a lower limit and marked with an asterisk.}

\item{S$_{\rm tot}$ - The total flux of the cloud determined by summing over the ROI in the Galactic background subtracted data.  Since the flux is calculated within the ROI identified in the convolved cube, for particularly compact objects it is typically 1-2 times lower than the total integrated flux.}

\item{Population - The classification of the cloud using the scheme detailed in \S \ref{sec:results}: 
\begin{itemize}
\item[HVC -]{(high velocity cloud) $|$\vlsr$|>90$ km s$^{-1}$ and near a known HVC complex}
\item[GC -]{(galaxy candidate) $|$\vlsr$|>90$ km s$^{-1}$ and far from known HVC complex}
\item[CLVC -]{(cold low-velocity cloud) $|$\vlsr$|<90$ km s$^{-1}$ and $\Delta V < $ 15 km s$^{-1}$}
\item[WLVC -]{(warm low-velocity cloud) $|$\vlsr$|<90$ km s$^{-1}$, $\Delta V > $ 15 km s$^{-1}$, and not Q3}
\item[Q3 -]{(warm, positive low-velocity cloud in the third Galactic quadrant) \\ $90 >$\vlsr$> 0$ km s$^{-1}$, $\Delta V > $ 15 km s$^{-1}$,
and $270>l>180$} 
\end{itemize}}

\end{itemize}

\section{Results}
\label{sec:results}

The conditions for a cloud to be included in this catalog are an angular size smaller than 20\arcmin~and separation from other HI emission
both spatially and in velocity. Without distance constraints, the clouds identified could have masses that range over many orders of magnitude.
From $10^{-3}-10^{-1}$ M$_\odot$ stellar outflows \citep{matthews11,matthews12}, to $10^1-10^2$  M$_\odot$ recooling 'cloudlets' \citep{heitsch09}, 
to $10^2-10^4$  M$_\odot$ clouds at the disc-halo interface \citep{ford08, ford10}, to  $10^5$  M$_\odot$ gas-rich dwarf galaxies \citep{ryanweber08}, to $>10^6$  M$_\odot$
minihalos \citep{giovanelli10}, observations and theories have shown that compact neutral hydrogen clouds are a heterogenous group in need of 
multi-wavelength analysis. With this in mind, we have examined the distribution of observable properties and separated the catalog into five
populations based on velocity, linewidth, and position. We recognize that this grouping is inexact and that the populations are continuous across
several of the boundaries, but we find that there are distinct differences between the populations. In this section, we discuss our 
cloud classifications and describe the observed properties of each population. In \S \ref{sec:discussion} we discuss how each population
relates to previous observations and models.

Our first step in separating the clouds into populations is to split the catalog into low- and high-velocity clouds at $|$\vlsr$| = 90$~\kms~as used in previous studies\citep{wakker97}. We chose not to use the deviation velocity
(V$_{\rm dev}$) introduced by \cite{wakker91_1} because the GALFA-HI DR1 coverage includes mostly high absolute Galactic latitudes and the 
Galactic anti-center where \vlsr $\simeq$ V$_{\rm dev}$. The greatest difference between \vlsr~and V$_{\rm dev}$ in the DR1 region is only 16~\kms.
 The high-velocity clouds are further separated into two populations by position and velocity: high-velocity clouds
(HVC) that are near known HVC complexes as described in \S \ref{sec:results_hvcs} and galaxy candidates (GC) that are far from known HVC
complexes. To motivate separating the low-velocity clouds, we note that 83\% of the HVCs have $\Delta V > 15$\kms~(see Figure \ref{fig:histo2}).
This is expected for partially-ionized warm neutral medium \citep[WNM]{wolfire95,wolfire03} and has been observed previously \citep[e.g.][]{putman02,hsu11,deheij02,kalberla06}. We use this observed boundary to separate the LVCs
into cold and warm populations. The cold low-velocity clouds (CLVC) have $|$\vlsr$|<90$ \kms~and $\Delta V < $ 15~\kms.
The final two populations are both warm low-velocity clouds. The warm, low-positive velocity clouds in the third Galactic quadrant ($90 >$\vlsr$> 0$ km s$^{-1}$, $\Delta V > $ 15 km s$^{-1}$, and $270>l>180$; Q3) are considered a different population from the rest of the warm low-velocity clouds (WLVC) due to their distinct velocity distribution (see \S \ref{sec:warmq3results}).

\subsection{High-Velocity Clouds}
\label{sec:results_hvcs}
We find 692 clouds in our catalog that meet the standard definition of high-velocity clouds (HVCs) of $|$\vlsr $| >$ 90 \kms~and are not in the galaxy candidates below. 
HVCs are thought to be clouds of gas that reside in the halo of our Galaxy, ranging in distance from a few kpc to at least 50 kpc for those associated 
with the Magellanic System \citep{wakker97,wakker04}. HVCs are typically found in groupings in position-velocity space called complexes. The HVCs detected in this catalog are very small compared to the scale of complexes which often span tens of degrees, and are called ultra-compact (size $< 20\arcmin$) HVCs (UCHVCs)\citep{bruns04}. 

To determine whether our UCHVCs are associated with known complexes we use the updated Wakker \& van Woerden (1991\nocite{wakker91}; WvW) catalog of HVCs. This catalog has the velocities, fluxes, and positions of 616 HVCs across the entire sky. Details on this sample of clouds can be found in Wakker (2004\nocite{wakker04}). To determine whether the 692 UCHVCs cataloged here are associated with known complexes of HVCs, we determine the ``distance'' in position-velocity space between a given catalog UCHVC and a HVC in the WvW catalog. We parameterize distance using the relationship from \cite{peek08}:

\begin{equation}\label{eq:deqn}
D = \sqrt{\Theta^2 + f^2\left(\delta v\right)^2}
\end{equation}

\noindent where $\Theta$ is the angular distance, $\delta v$ is the difference in velocity, and $f$ is a conversion factor Peek et al. set equal to $0.5$\degr/\kms. 
By analyzing simulations of HVCs, Peek et al. found that clouds from the same complex typically have $D<25\degr$. Figure \ref{fig:ra_v_hvc} is a position-velocity diagram where we have colored the HVCs by which complex they are associated with. Most notable are the large number of clouds associated with the tip of the Magellanic Stream (MS), the anti-center complex (AC), and the WA, WB and WC complexes. 
Figure \ref{fig:aitoff} shows the cloud positions  and GALFA-HI DR1 coverage in Galactic coordinates with outlines of the HVC complexes.
We find that the majority of our HVCs have $D<$~9\degr~to the nearest WvW HVC, with only 35 HVCs having $D > 25$\degr. 
We separate these isolated clouds into a sub-population of galaxy candidates discussed in the following section.
None of the $-400<$ \vlsr $< -90$ \kms~clouds have $D>$ 35\degr , leaving 86\% of the negative HVC parameter space completely empty. 

\begin{figure}
\begin{center}
\includegraphics[scale=.85,angle=0]{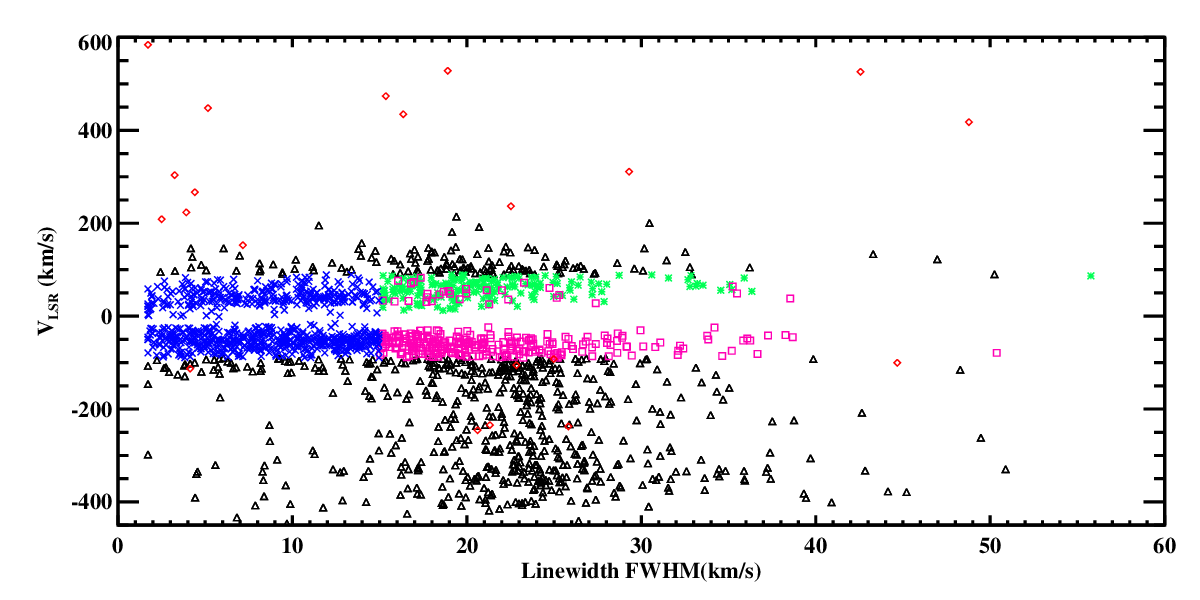}
\caption{Linewidth and \vlsr distributions for all clouds. The differently colored symbols correspond to 
the five populations - HVCs (black triangles), galaxy candidates (red diamonds), cold LVCs (blue Xs), warm LVCs (pink open squares), and warm Q3 LVCs (green filled squares). This plot best illustrates where the populations are separated. See Figure \ref{fig:histo2} for the velocity and linewidth
distribution for each population.
\label{fig:line_v_plot}}
\end{center}
\end{figure}

\begin{figure*}
\begin{center}
\includegraphics[scale=.8, angle=0]{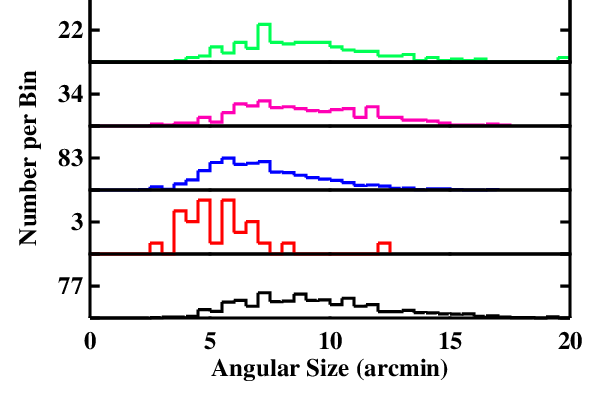}
\includegraphics[scale=.8, angle=0]{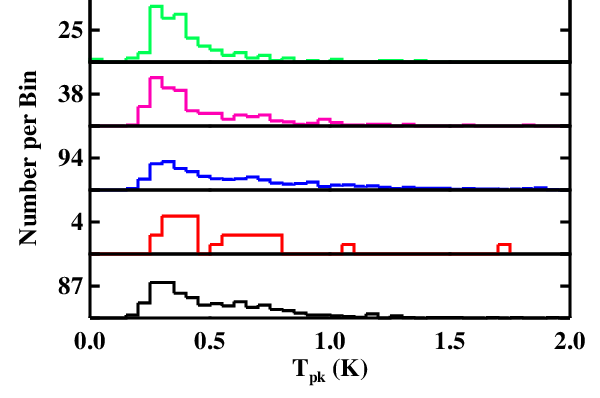}
\includegraphics[scale=.8, angle=0]{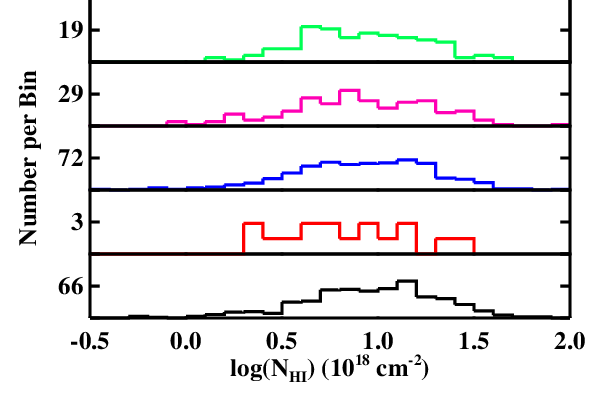}
\includegraphics[scale=.8, angle=0]{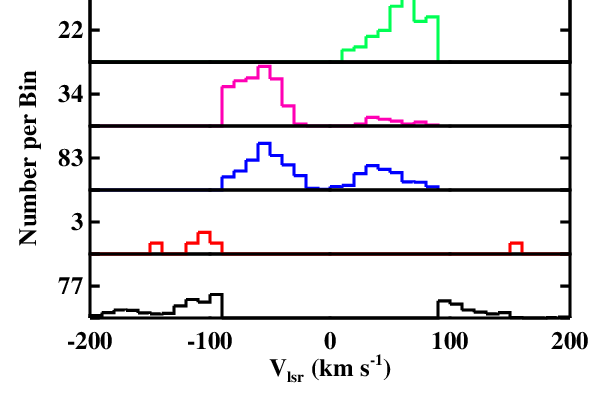}
\includegraphics[scale=.8, angle=0]{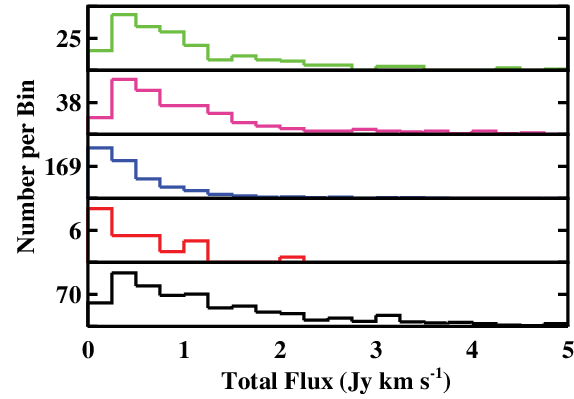}
\includegraphics[scale=.8, angle=0]{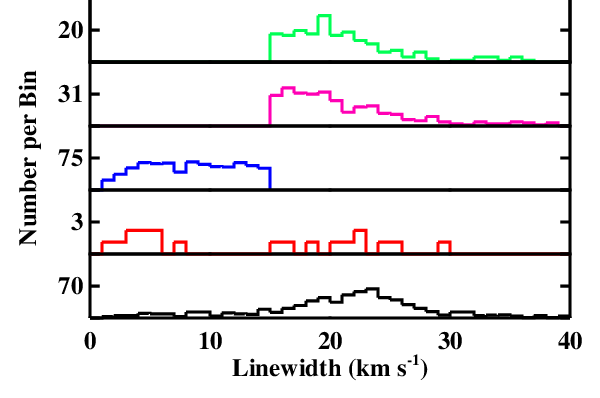}
\caption{The distributions of angular size, peak brightness temperatures (T$_{\rm pk}$), column density, LSR velocity, total flux and linewidth for all cataloged clouds, separated by population - HVCs (black, lower pane), galaxy candidate (red, second pane), cold LVCs (blue, center pane), warm LVCs (pink, fourth pane), and warm Q3 LVCs (green, upper pane). To aid in comparing the different populations, each histogram is scaled so that the fractional number of clouds per bin is consistent for each plot. The y-axis values are the mid-values for each histogram in number of clouds. \label{fig:histo2}}
\end{center}
\end{figure*}

The HVCs are shown in black in Figures~\ref{fig:ra_dec}, \ref{fig:line_v_plot}, and \ref{fig:histo2}.  Figure \ref{fig:line_v_plot} shows the linewidth distribution for all the clouds in the catalog. We do not separate the HVCs by linewidth and note that 83\% have $\Delta V >15$ \kms. The measured properties of the clouds are plotted in Figure \ref{fig:histo2}. The size distribution of the HVCs is peaked near 8\arcmin~with a tail to our limit of 20\arcmin. 
The number of HVCs increases at lower peak brightness temperatures, as we see for all the populations. There is a sharp decline in clouds below T$_{Pk}$ = 0.3 K, which corresponds to 
$\sim3 \sigma$ in the unconvolved data cubes. The column density distribution peaks near $\sim10^{19}$ cm$^{-2}$ and extends an order of magnitude higher and lower.
There is a sharp break in the velocity distribution of the HVCs at 90 \kms~where we define the boundary of HVCs and LVCs, indicating some of the LVCs may be low velocity extensions of HVCs. We discuss this further in \S\ref{sec:warmq3discussion}.  At the upper ends, the UCHVCs extend to just below -400 \kms~ (representative of gas at the tail of the Magellanic Stream) and generally under 200 \kms~on the positive velocity side.

\subsubsection{Galaxy Candidates}

\label{sec:galaxy}
\begin{figure}
\begin{center}
\includegraphics[scale=.93, angle=0]{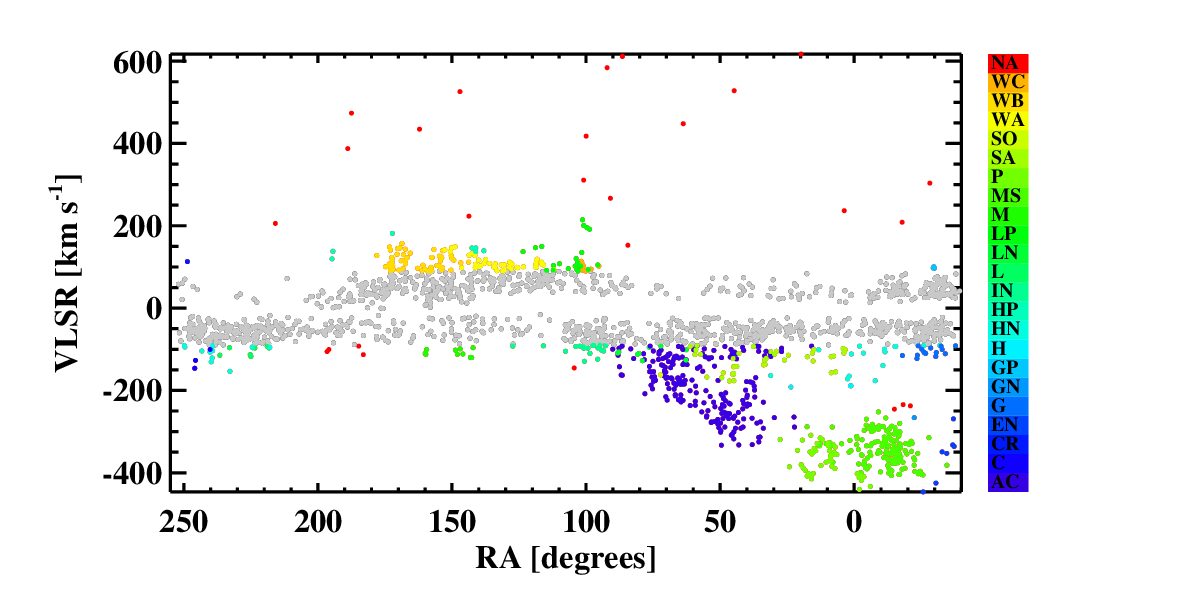}
\caption{We show the distribution of all cataloged clouds in RA vs. \vlsr, highlighting the HVCs. Clouds the do not meet the simple HVC criterion of $|$\vlsr$| > 90$ \kms are marked in gray. Clouds with a $D$  (see Equation \ref{eq:deqn}) to the closest WvW catalog HVC less than 25\degr are marked in the color denoting that nearest complex or grouping in the color bar at right. Clouds marked in red have $D$ greater than 25\degr~to any HVC in the WvW catalog. Note the strong asymmetry across \vlsr$ = 0$\kms in unassociated clouds.\label{fig:ra_v_hvc}}
\end{center}
\end{figure}

There are 27 catalog members which lie at $|$\vlsr$|$ $>90$~\kms~and
are not correlated with known galaxies or HVC complexes.  These are
represented by red in Figures~\ref{fig:line_v_plot},
\ref{fig:ra_v_hvc}, and \ref{fig:histo2}.  Of these 27, 8 are at
negative velocities and although they lie outside the threshold for
association discussed in \S\ref{sec:results_hvcs}, they do appear
to be extensions in position-velocity space of known HVC complexes.
The remaining 19 positive velocity galaxy candidates range in
velocity from 153~\kms~to 617~\kms, with a median value of 418~\kms. A plot of RA vs. \vlsr
~is shown in Figure \ref{fig:ra_v_hvc}, where this population is shown in red.

It is likely that many of these high velocity catalog members
represent undiscovered galaxies, given their clear bias towards
recessional velocities.  Supporting this hypothesis is their small
spatial extent, which has a median size of 5\arcmin.
While these objects do not have peak column densities or brightnesses
similar to the 58 known galaxies in our sample (average N$_{\rm HI}
= 1.69\times 10^{20}$ cm$^{-2}$ and T$_{B} =1.5$K), this may be
expected for low surface brightness dwarf galaxies that have not yet
been discovered optically. Many of these are the same detections
discussed by \cite{giovanelli10} as possible mini halos.

We compared the sample of 27 potential galaxies to the complementary
Arecibo Legacy Fast ALFA Survey (ALFALFA) catalog (Haynes et al.
2011\nocite{haynes11}).  Much of the DR1 data was taken commensally
with this survey.
Eight of the 27 GALFA-HI galaxy candidates are in the ALFALFA coverage
region, and 6 of these are in the ALFALFA catalog.  For this sample of
six, Haynes et al. searched for optical counterparts and 
found 2 to lie outside the clean SDSS coverage
region, 2 to be of uncertain SDSS association (e.g. displaced in
redshift, or near objects without known redshifts), and 2 to have no
optical counterpart at all in SDSS.

\subsection{Cold Low-Velocity Clouds}
\label{sec:cold}
We identify 1245 clouds with $|$\vlsr$| < 90$ km s$^{-1}$. 
For purely thermal broadening, the temperature and linewidth of a cloud are related by
\begin{equation}
T=21.8{\rm K} \left(\frac{\Delta V}{\rm km~s^{-1}}\right)^2. 
\end{equation}
\noindent The upper limit on the temperature of a cloud with $\Delta {\rm V} = 15$~\kms~is $\sim$5000K which is a typical lower bound on the temperature of the partially-ionized warm neutral medium (WNM) \citep{wolfire03}.
We separate these low and intermediate velocity clouds into 750 CNM-dominated cold clouds with average linewidths less than 15 \kms~and 
495 WNM-dominated warm clouds with average linewidths greater than 15 \kms. 
As Figure \ref{fig:line_v_plot} shows, the majority of the HVCs in the catalog are thus classified as warm, as is expected for clouds embedded in the hot halo \citep{wolfire03}.

\begin{figure}
\begin{center}
\includegraphics[scale=.6, angle=0]{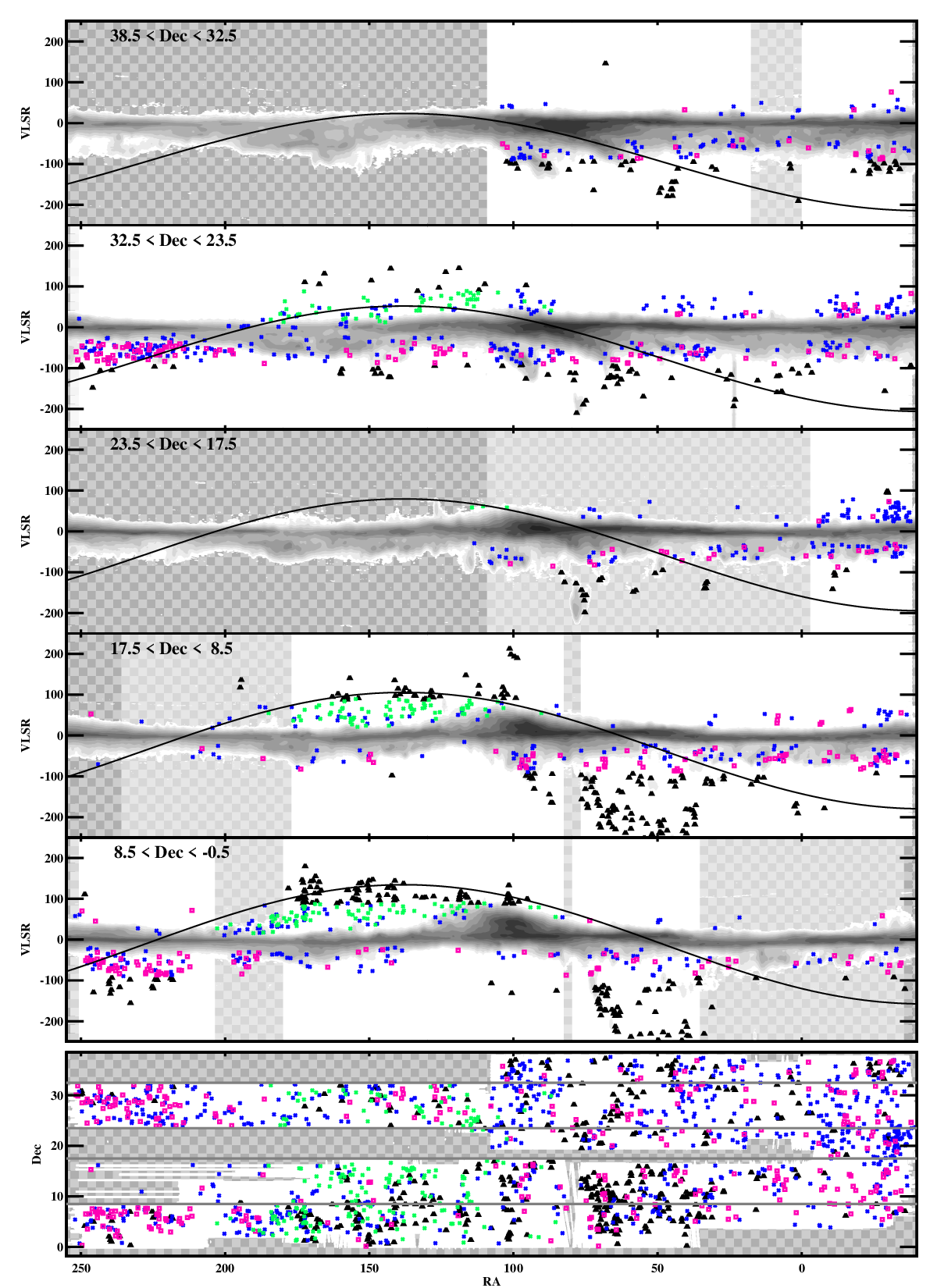}
\caption{Position-position-velocity distribution of clouds with \vlsr$< \pm 250$\kms. The cloud populations are colored as in Figure \ref{fig:line_v_plot}. {\it Bottom Panel}: Position-position plot showing the GALFA-HI DR1 coverage. The hashed region is not included in DR1. The grey horizontal lines indicate the breaks between the declination regions used in the upper panels.
{\it Upper Panels}: Position-velocity plots showing clouds in the declination range indicated. The shading indicates the relative amount of observations in the region. The greyscale image is the LAB survey \citep{kalberla05} integrated along the declination range indicated.
The curved line follows the velocities corresponding to 
\vgsr$=0$~\kms, where a cloud at rest compared to Galactic rotation would lie.
\label{fig:ra_dec}}
\end{center}
\end{figure}

The cold LVCs are shown in blue in Figures~\ref{fig:ra_dec}, \ref{fig:aitoff}, \ref{fig:line_v_plot}, and \ref{fig:histo2}.  
From the histograms in Figure \ref{fig:histo2}, we see that the cold low-velocity clouds are relatively evenly distributed in velocity with 63\% having negative velocities. The number of cold LVCs also decreases significantly toward $\pm90$~\kms, unlike the warm LVCs (as discussed below).  Spatially, there appears to be an increase in density at $\alpha, \delta$ = 330\deg, 20\deg~($l,b \sim$ 77\deg, -27\deg) and an underdensity at positive velocities for $\alpha = 200-250$\deg~(general direction of the Galactic center and high positive Galactic latitude) (Figure~\ref{fig:ra_dec}). The cold LVCs are typically smaller than the warm LVCs and HVCs with a  peak centered at 6\arcmin~(near our resolution limit), though there is also a tail extending to 20\arcmin. 
This indicates that some of these clouds are unresolved.  The linewidths also extend to the lower limit of our search space ($\sim$3~\kms), indicating there are even colder clouds yet to be found.
The cold clouds have the shallowest peak brightness distribution with the 95 percentile at 1.8K. Similar to the other populations, there is a sharp decline in the number of clouds with peak brightnesses below 0.3K, effectively our sensitivity cutoff.
The steepness of this decline indicates that there may be many more clouds to be found with deeper observations. 
Even though the cold LVCs have higher brightness temperatures, they have
lower total flux and column density values due to narrower linewidths and smaller sizes.

\subsection{Warm Low-Velocity Clouds}
\label{sec:warmresults}
Of the cataloged clouds, 495 have low velocities with $|$\vlsr$| < 90$ km s$^{-1}$ and linewidths of $\Delta V > 15$ km s$^{-1}$.  As described in \S\ref{sec:cold}, this corresponds to temperatures greater than 5000K. 
We identify a sub-population of warm LVCs with positive velocities positioned in Q3 that we discuss in \S\ref{sec:warmq3results} and \S\ref{sec:warmq3discussion}.
With the removal of the clouds in Q3, 302 clouds remain, 90\% of which have \vlsr $<0$~\kms (magenta in the figures). In Figure \ref{fig:histo2}, we see that the velocity distribution of the warm LVCs continues to -90 \kms~where we separate them from the HVCs, indicating that these two populations are most likely blended in velocity or even partially part of the same population. There is a sharp drop in clouds approaching 0 \kms~due to confusion with larger structures of low velocity gas.
The angular size distribution is 
similar to the HVCs with a median value of 9\arcmin~and extending to the limits of the catalog. The total flux and column density distributions are also similar to the HVC distributions. We observe a steeper peak brightness distribution than for the cold LVCs with 95\% of warm LVCs below 1.0K.  The velocity widths show a sharp cut-off at our break at 15 \kms~indicating there is most likely a continuum of clouds in temperature.  There appears to be overdensities of warm clouds located at $200\degr$-$250\degr$ and $330\degr$ in RA, the locations of IVC complexes K and the PP Arch, respectively \citep{wakker01}.

\subsubsection{Warm, Positive-Velocity Q3 Clouds}\label{sec:warmq3results}
The vast majority of warm clouds with $90 >$\vlsr$> 0$ km s$^{-1}$ are located in the third Galactic quadrant (RA=5-14hr; green in the figures).
Since our coverage does not include the fourth quadrant (see Figure \ref{fig:aitoff} for the GALFA-HI DR1 coverage in Galactic coordinates), the third quadrant is the only area of the sky where infalling (or static) clouds may have 
positive LSR velocities due to Galactic rotation. 
These clouds are below the traditional velocity cut for HVCs, but otherwise have similar properties to the HVCs. 
In Figure \ref{fig:ra_dec}, we plot the velocities that correspond to \vgsr=0~\kms~to illustrate the effect of Galactic rotation. 
In the Galactic Standard of Rest (GSR), these clouds have velocities near or below zero, as do the HVCs associated with the WA, WB, and WC Complexes. This suggests that the positive velocities in the LSR frame are a product of Galactic rotation, and that these clouds 
have low or negative velocities relative to the Galactic plane. In position, the warm Q3 clouds appear to be an extension of the IV-WA complex \citep{wakker01} which is located at similar Galactic latitudes but in the fourth Galactic quadrant.

From Figure~\ref{fig:histo2}, the warm Q3 clouds have size, total flux, and column density distributions similar to the rest of the warm LVCs and the HVCs. The linewidth distribution is similar to the warm clouds with a sharp cut at 15~\kms~and extending beyond 30~\kms. The warm Q3 clouds are the weakest population in brightness temperature with a 95 percentile of 0.8~K. The velocity distribution shows that the decrease in the number of clouds toward
\vlsr =0~\kms~begins at a higher velocity than the rest of the warm or cold cloud populations.  This may be partially due to only including a limited range of positions for this population in the direction of positive Galactic rotation; however, there does appear to be a larger gap between the warm Q3 clouds and
the Galactic disc in Figure \ref{fig:ra_dec} than the rest  of the warm LVC population.

\section{Discussion}
\label{sec:discussion}

In this section we discuss how the different populations of clouds we have defined in \S\ref{sec:results} fit into current theories and how they relate to previous observations. 

\subsection{UCHVCs}

UCHVCs are especially of interest because it has been suggested that small HVCs may represent physically large structures at megaparsec distances,
much farther than the classical complexes\citep{braun99, giovanelli10}. It is very hard to obtain direct distance information for such small clouds, given the unlikely chance of an overlap with a halo star;  however, if the UCHVCs reside near larger complexes in position-velocity space, we expect they are associated with these relatively nearby structures, rather than being independent clouds at much larger distances. 

As discussed in \S\ref{sec:results_hvcs}, all but a few UCHVCs have $D<25\degr$ to an HVC in the WvW catalog. At negative velocities, this leaves 86\% of the available position-velocity space with $-400 <$\vlsr$<-90$~\kms~empty. 
The presence of UCHVCs found only near other known HVC complexes is consistent with the bulk of the Galactic hot halo not cooling due to linear thermal instabilities \citep{binney09, joung11}.  If the bulk of the halo were thermally unstable to linear perturbations, we would expect small neutral condensations to appear all over phase space, rather than only concentrated toward known, larger HVCs. This result instead supports the idea that some HVCs may be seeded by larger perturbations in the halo, such as filamentary streams of gas impinging on the Galaxy from the intergalactic medium or satellite galaxies\citep{keres09, joung11}.

If UCHVCs are indeed associated with larger clouds, they are most likely at similar distances. The median UCHVC properties indicate a mass of $\sim17$~M$_{\odot}$ and a physical size of $\sim10$~pc at a distance of 10~kpc and a mass of $\sim1700$~M$_{\odot}$ and a physical size of $\sim100$~pc at 100~kpc.  These distances were chosen based on the distance constraints to many HVC complexes \citep{thom06,wakker01, wakker08}, and the likely distance of the tail of the Magellanic Stream given simulations \citep{besla10, connors06}.  The mass of these clouds indicate they are not likely to survive the trip to the Galactic disk and will instead become part of the multi-phase Galactic halo \citep{heitsch09}.  The exception would be if some of the UCHVCs are embedded within extended, highly ionized shells.

\subsubsection{Warm, Positive-Velocity Q3 Clouds}\label{sec:warmq3discussion}

In \S\ref{sec:warmq3results} we state that the vast majority of the positive velocity warm low velocity clouds ($0<$\vlsr$<90$~\kms) are located in the third Galactic quadrant and are likely to be associated with the W HVC complexes. In Figure \ref{fig:ra_dec} we have plotted the \vlsr~
velocities corresponding to \vgsr$=0$~\kms~as a curved line. Clouds on this line could be at rest relative to Galactic rotation. The majority of the positive-velocity warm clouds in Q3 are bounded by the \vgsr$=0$~\kms~line indicating these clouds may be infalling clouds with positive LSR velocities due to Galactic rotation.

The HVC complexes WA, WB, and WC (e.g. WvW) are in the same region of sky as the warm Q3 clouds, and if we extend the associations with HVCs from \S\ref{sec:results_hvcs} below 90 \kms, many of the Q3 clouds would be considered part of these complexes. \cite{thom06} measured a direct distance constraint to clouds in the W complexes and found a distance of $\sim9$ kpc.  This results in a z-height of 7 kpc and galactocentric distance of 12 kpc. 
We consider three scenarios for a cloud at this distance at ($\alpha$,$\delta$)=(150\deg,15\deg):  a non-rotating cloud, a co-rotating cloud, and a co-rotating cloud with a vertical lag. For a non-rotating cloud (\vgsr=0), we would observe \vlsr $\simeq95$~\kms~due to the rotation of the Galaxy. If the cloud were co-rotating at 220~\kms, we would observe \vlsr$\simeq10$~\kms. If the cloud were co-rotating, but with a vertical lag of 20~\kms~kpc$^{-1}$, we would observe \vlsr $\simeq65$~\kms. All of these calculations are without any vertical infall which would lower the observed \vlsr.
The velocities we observe for the warm Q3 clouds are consistent with non-rotating clouds with infalling velocities of $\sim$10-70~\kms~or co-rotating clouds in a lagging halo.
The warm Q3 clouds may fall under the category of the `low-velocity halo clouds' \citep{peek09}, and may also represent a bridge connecting the HVC W complexes to the Galactic disc.

\subsection{Cold Low-Velocity Clouds}
The cold, low-velocity clouds are certainly associated with the disc of the Galaxy. There are greater numbers of these clouds at lower velocities and 
there is no apparent relationship to the GSR frame like for the warm Q3 clouds (\S\ref{sec:warmq3discussion}). The cold LVCs may be related to the clouds
studied by \cite{ford08, ford10} who identified a population of discrete, cold clouds 600-1000pc out of the plane of the Galaxy that are co-rotating
with the disc. Their measurements were made on a set of clouds with Galactic latitude less than $20\degr$ and at a distance of $\sim$8kpc in the first and third Galactic quadrants.
Ford et al. determined distances by selecting clouds above the tangent point velocity, which is the maximum velocity allowed by
Galactic rotation toward sightlines of the inner Galactic plane.
The GALFA-HI DR1 does not include the inner Galactic plane so we cannot use the tangent point method to determine distances.  
Much of this catalog is at high Galactic latitudes where 
the Ford clouds would be closer ($<1$kpc) and therefore appear substantially larger, up to two degrees, though many of their clouds were unresolved. 
The masses of the Ford sample range from $\sim100$ to $\sim5000$ M$_\odot$, while the cold LVCs from our catalog would have masses of $\sim0.15$ M$_\odot$ at a similar vertical position.
High latitude analogs of the Ford clouds
would not be included in this catalog due our size limit of 20\arcmin, while analogs to the clouds in this catalog would not have been detected by Ford et al. due to the small sizes and masses. 
However, the clouds in the Ford sample have a median linewidth of 10.6 \kms, while our cold clouds have a median linewidth of  8.4 \kms, 
indicating both populations have typical temperatures less than 5000K. The cold LVCs in this catalog are not physically the same as the Ford clouds, 
but may have similar temperature, vertical structure, and kinematics.

Two possible origins of the cold LVCs are a Galactic fountain process or recooling disc-halo interface gas. In a Galactic fountain model, gas is expelled out of the plane
of the Galaxy by some energetic process and then rains back down out of the halo \citep{shapiro76}. Whether the ejected gas is localized to regions of star formation, or rapidly blends with existing warm/hot lower halo gas and condenses later is still unknown.   This will be investigated further with models and a larger population of GALFA-HI clouds in the future. 
Regarding recooling gas, \cite{heitsch09} observed small clouds forming from the warm remnants of disrupted HVCs in their simulations as the clouds became buoyant close to the disk.  
The large number of positive-velocity cold LVCs requires some process to drive them away from the disc, be it the buoyancy of accreting gas or feedback from star formation in the disc.

\subsection{Warm Low-Velocity Clouds}

After classifying the positive-velocity warm clouds in the third Galactic quadrant as possible low-velocity halo clouds, we see that 90\% of the remaining warm 
low-velocity clouds have negative velocities. This agrees well with previous studies of intermediate velocity ($|$\vlsr$|$ $=40-90$ \kms) clouds 
that found most of this gas is at negative velocities and has a vertical distance ranging from 0.5 to 3 kpc 
(See \cite{albert04} for an extensive discussion). The warm low-velocity clouds detected here may be associated with IVCs. 
If these clouds are within 3~kpc of the disc, they would have masses of $\sim0.1-10$~M$_\odot$.
The origin of the clouds is uncertain but the higher metallicities for some of the large IVCs \citep{wakker01} suggests a relation to processes within the Galactic disc. As shown in Figure \ref{fig:histo2}, the velocity and line width distributions are continuous across the 90 \kms~ and 15 \kms~ boundaries, respectively. This suggests that the warm LVCs may be partially low-velocity HVCs and partially warm or turbulent cold LVCs. This is different from the cold LVCs that appear separate from the HVCs in velocity. From this reasoning, some of the warm LVCs are most likely the remnants of disrupted HVCs, while other warm LVCs may have come from, or have yet to become, cold LVCs.  In any case, the negative velocities of these clouds indicate they will be future star formation fuel.  

\section{Summary}

Using a custom cloud detection algorithm, we have identified 1964 compact ($<20$\arcmin) neutral hydrogen clouds with velocities typically between $|$\vlsr$|$ $= 20 - 450$ \kms, linewidths of $2.5-35$ \kms, and column densities ranging from $1-35 \times 10^{18}$ cm$^{-2}$. 
At high velocities, we observe the vast majority of clouds are associated with previously known HVC complexes. 
At low velocities, we observe populations of cold and warm clouds that appear associated with the disc of the Galaxy, as well as a group of warm clouds at positive velocities that may be low-velocity halo clouds associated with HVC complexes WA, WB, and WC.  The catalog extends down to our limits 
spatially, spectrally, and in sensitivity, indicating that while we have discovered a large number of new small structures, there are still smaller, colder and fainter clouds to be detected.

Although our search criterion was simply for clouds smaller than 20\arcmin~in the GALFA-HI DR1 area, we detected clouds in vastly different environments. From potentially undiscovered galaxies, to HVCs associated with the Magellanic Stream at 100~kpc, to complexes of clouds at $<10$kpc, to clouds associated with the Galactic disc, we find differences in the distribution of clouds indicating a range of origins.
While the HVCs associated with complexes may be sheared off the larger clouds by instabilities, the LVCs could have been ejected from the disc and some may represent a link to HVCs. The distances to the clouds in this catalog may cover several orders of magnitude, so the masses may range from less than a Solar mass for clouds within the Galactic disc, to greater than $10^4$ M$_\odot$ for HVCs at the tip of the Magellanic Stream.

This catalog is ripe for multi-wavelength analysis and follow up observations. 
Finding optical counterparts to the galaxy candidates will provide distances and masses, and work is already underway to identify potential new ultrafaint dwarf galaxies with the catalog (Grcevich et al., in prep.). 
Infrared data will allow us to determine the dust content of these clouds and may disentangle the halo clouds from the disc clouds.  We are also pursuing Herschel spectral line observations to investigate the possibility of cooling within the clouds, and will look for any fortunate overlap with stars to provide direct distance and potentially metallicity constraints.
Finally, future GALFA-HI observations combined with Galaxy models will further constrain the origin and role of the smallest and coldest population of individual HI clouds yet found.

\acknowledgements{We acknowledge the staff at Arecibo Observatory for their ongoing support and our commensal partners in the ALFALFA and GALFACTS teams for their help taking the observations. DRS thanks M.-M. Mac Low, F. Paerels, and M. Ag\"{u}eros for their comments and insight. JEGP was supported by HST-HF-51295.01A, provided by NASA through a Hubble Fellowship grant from STScI, which is operated by AURA under NASA contract NAS5-26555. This work was supported in part by NSF Collaborative Research grants AST-0707679, 0709347, 0917810, NSF CAREER grant 094059, NASA Herschel grant 1369759, and the Luce Foundation.}

\bibliographystyle{apj}
\bibliography{cloudcatalog}

\clearpage

\begin{landscape}
\begin{deluxetable}{cccccccccccccc}
\tabletypesize{\scriptsize}
\setlength{\tabcolsep}{0.05in}
\tablewidth{8in}
\tablecaption{DR1 GALFA-HI Cloud Catalog}
\tablehead{ \colhead{Cloud} & \colhead{RA} & \colhead{DEC} & \colhead{l} & \colhead{b} & \colhead{$\sigma_{\rm Max}$} &\colhead{Size} & \colhead{V$_{\rm LSR}$} & \colhead{V$_{\rm GSR}$}  & \colhead{$\Delta V$} & \colhead{$T^{Pk}_b$} & \colhead{N$_{HI}$} & \colhead{S$_{tot}$}
& \colhead{Population}\\
ID & (h:m:s) & (d:m) & (deg) & (deg) &  & (arcmin) & (km s$^{-1}$) & (km s$^{-1}$) & (km s$^{-1}$) & (K) & ($10^{18}$ cm$^{-2}$) & (Jy km s$^{-1}$) }
\startdata
001.4+35.7+030 & 00:05:29 & 35:40 &  112.54 &  -26.27 &    16.4 &  6.5 &   30.98 &  213.19 &    8.42 &    1.29 & 16 &    0.860 & CLVC\\
001.9+16.3-350 & 00:07:37 & 16:17 &  107.91 &  -45.30 &    14.8 &  8.1 & -346.88 & -199.62 &   37.42 &    0.45 & 32 &    2.093 & HVC\\
096.3+35.2-079 & 06:25:13 & 35:09 &  178.59 &   10.34 &    22.7 & 12.2 &  -80.00 &  -74.66 &    4.26 &    1.69 & 10 &    2.140 & CLVC\\
109.7+14.7+063 & 07:18:49 & 14:39 &  202.74 &   12.60 &    50.7 & 19.5 &   64.70 &  -18.31 &   21.08 &    0.83 & 33 &    13.783 & Q3\\
143.1+27.1-119 & 09:32:29 & 27:06 &  201.03 &   46.04 &    13.9 &  7.0 & -119.42 & -174.21 &    4.78 &    0.48 & 3 &    0.264 & HVC\\
175.8+32.2-070 & 11:43:21 & 32:14 &  189.75 &   74.38 &     8.3 &  4.7 &  -70.12 &  -80.15 &    3.10 &    0.34 & 2 &    0.077 & CLVC\\
\enddata
\label{tab:cattab}
\tablecomments{An excerpt from the full catalog table correspond to the clouds in Figure \ref{fig:cat}. The velocity in the cloud ID is determined from the peak position in the convolved cube and varies slightly from \vlsr~which is the velocity from the Gaussian fit. The full table is available at http://sites.google.com/site/galfahiccc.\\ *Average column density is reported where the fit through the maximum of the cloud fails to converge.  }
\end{deluxetable}
\end{landscape}

\clearpage

\begin{deluxetable}{lccccc}

  \tablecaption{Median Properties of Cloud Populations \label{tab:median_table}}
  \tablehead{ \colhead{Parameter} & \colhead{HVC} & \colhead{Gal} & \colhead{Cold} & \colhead{Warm} & \colhead{Q3}}
  \startdata
  V$_{\rm LSR}$(km s$^{-1}$) & -166 & 266 & -38 & -59 & 61\\
  $T_{\mathrm{pk}}$ (K) & 0.4 & 0.4 & 0.5 & 0.4 & 0.4 \\
  $\Delta v$ (km~s$^{-1}$) & $22.0$ & $21.3$ & $8.4$ &$ 20.0$  & $20.0$\\
  $N_{HI}$ ($\times 10^{18}$ cm$^{-2}$) & $11$ & $8$ & $6$ & 9  & 9 \\
  Angular Size ($'$) & $9$ & $5$ & $7$ & 9 & 9 \\
   \enddata
   \label{table:median}
\end{deluxetable}

\begin{deluxetable}{lccccc}

  \tablecaption{90\% Range of Properties of Cloud Populations \label{tab:ninety_table}}
  \tablehead{ \colhead{Parameter} & \colhead{HVC} & \colhead{Gal} & \colhead{Cold} & \colhead{Warm} & \colhead{Q3}}
  \startdata
  V$_{\rm LSR}$(km s$^{-1}$) & $-389 \rightarrow 126$ & $-237 \rightarrow 612$ & $-79 \rightarrow 64$ & $-89 \rightarrow -48$ & $21 \rightarrow 86$ \\
  $T_{\mathrm{pk}}$ (K) & $0.2 \rightarrow 1.1$ & $0.3 \rightarrow 1.1$& $0.3 \rightarrow 1.8$  & $0.2 \rightarrow 1.0$ & $0.2 \rightarrow 0.8$\\
  $\Delta v$ (km~s$^{-1}$) & $6.3 \rightarrow 34.3$ & $2.5 \rightarrow 69.8$ & $2.5 \rightarrow 14.2$  & $15.4 \rightarrow 33.8$ & $15.4 \rightarrow 32.8$\\
  $N_{HI}$ ($\times 10^{18}$ cm$^{-2}$) & $2 \rightarrow 34$ & $2 \rightarrow 70$ & $1 \rightarrow 19$  & $2 \rightarrow 28$ & $3 \rightarrow 26$\\
  Angular Size ($'$) & $5 \rightarrow 15$ & $3 \rightarrow 8$ & $4 \rightarrow 12$ & $5 \rightarrow 14$ & $5 \rightarrow 14$\\
  \enddata
  \label{table:range}
\end{deluxetable}

\clearpage

%%%%%%%%%%%%%%%%%%%%%%%%%%%%%%%%%
%%%%%%%%%%%%%%%%%%%%%%%%%%%%%%%%%

\clearpage

\end{document}